\documentstyle[12pt,aasms4]{article}
\begin{document}

\title{Erratum: ``Intergalactic Photon  Spectra from the  Far IR to the  
UV Lyman Limit for $0  < z < 6$ and  the Optical Depth of the  Universe to High
Energy Gamma-Rays'' (ApJ 648, 774 [2006])}

\author{F.W.    Stecker}  \affil{NASA/Goddard  Space   Flight  Center}
\authoraddr{Greenbelt,     MD     20771;     Floyd.W.Stecker@nasa.gov}
\affil{Department of Physics  and Astronomy, University of California,
Los  Angeles}   \authoraddr{Los  Angeles,  CA90095-1547}  \author{M.A.
Malkan}  \affil{Department  of Physics  and  Astronomy, University  of
California,   Los  Angeles}  \authoraddr{Los   Angeles,  CA90095-1547;
malkan@astro.ucla.edu}
\author{S.T.   Scully}  \affil{Department  of Physics,  James  Madison
University} \authoraddr{Harrisonburg, VA 22807; scullyst@jmu.edu}

Table 1 in our paper had erroneous numbers for the coefficients fitting
the parametric form for the optical depth of the universe to
$\gamma$-rays, $\tau.$ The correct values for these parameters as
described in the original text are given in the table below for various
redshifts for the baseline model (upper row) and fast evolution (lower
row) for each individual redshift. The parametric approximation holds
for $10^{-2} < \tau < 10^2$ and $E_{\gamma} < \sim2$ TeV for all redshifts 
but also up to ~10 TeV for redshifts, $z \le 1$.

\begin{table}
\centerline{Table 1: Coefficients for the Baseline and Fast Evolution  Fits}
\begin{center}
\begin{tabular}{|cccccc|}  \hline \hline
$z$ & $A$ & $B$ & $C$ &  $D$ & $E$\\
0.03 & -0.020228 & 1.28458 & -29.1498 & 285.131 & -1024.64 \\ 
~&  -0.020753  &    1.31035   &  -29.6157   &   288.807  &   -1035.21 \\
0.117 & 0.010677 & -0.238895 & -1.004 & 54.1465 & -313.486 \\ 
~& 0.022352  &   -0.796354 &    8.95845 &    -24.8304  &   -79.0409 \\
0.2 & 0.0251369 & -0.932664 & 11.4876 & -45.9286 & -12.1116 \\
~&  0.0258699  &  -0.960562 &    11.8614  &   -47.9214   &  -8.90869 \\
0.5 & -0.0221285 & 1.31079 & -28.2156 & 264.368 & -914.546 \\ 
~&  0.0241367  &  -0.912879 &    11.7893 &    -54.9018   &   39.2521 \\
1.0 & -0.175348 & 8.42014 & -151.421 & 1209.13 & -3617.51 \\ 
~& -0.210116  &    10.0006 &    -178.308  &    1412.01  &   -4190.38 \\
2.0 & -0.311617 & 14.5034 & -252.81 & 1956.45 & -5671.36 \\ 
~& -0.397521  &    18.3389  &   -316.916  &    2431.84 &    -6991.04 \\
3.0 & -0.34995 & 16.0968 & -277.315 & 2121.16 & -6077.41 \\
~ & -0.344304 & 15.8698 & -273.942 & 2099.29 & -6025.38\\ 
5.0 & -0.321182 & 14.6436 & -250.109 & 1897.00 & -5390.55 \\
~&-0.28918 &  13.2673 & -227.968 &  1739.11 &  -4969.32 \\

\hline
\end{tabular}
\end{center}
\end{table}

\end{document}